\documentclass{emulateapj}

\newcommand\etal{{\it et al.~}}

\begin{document}

\title{AMBIPOLAR-DIFFUSION TIMESCALE, STAR-FORMATION TIMESCALE, AND THE AGES OF MOLECULAR CLOUDS:
IS THERE A DISCREPANCY ?} 

\author{Konstantinos Tassis \& Telemachos Ch. Mouschovias}

\affil{Departments of Physics and Astronomy \\
University of Illinois at Urbana-Champaign, 1002 W. Green Street, Urbana, IL 61801}

\begin{abstract}
We re-examine critically the estimates of the duration of different phases of star formation and the 
lifetimes of molecular clouds, based on the ages of T-Tauri stars, age spreads of stars in clusters, 
and statistics of pre-stellar cores. We show that all available observational data are
consistent with lifetimes of molecular clouds comparable to $\approx 10^7$ yr, as well as
with the predictions of the theory of self-initiated, ambipolar-diffusion--controlled 
star formation. We conclude that there exists no observational support 
for either ``young'' molecular clouds or ``rapid'' star formation.
\end{abstract}


\section{Introduction}

According to the theory of self-initiated, ambipolar-diffusion--controlled star formation, 
ambipolar diffusion is responsible for the fragmentation of molecular clouds and the formation 
of thermally and 
eventually magnetically supercritical protostellar fragments (or cores) in 
initially magnetically subcritical parent clouds (see reviews by Mouschovias 1987, 1996, and 
references therein).
The formation of supercritical fragments takes place on a timescale between $10^5$ and a few 
$\times \, 10^7$ yr, 
depending on the initial mass-to-flux ratio, and degree of ionization 
of the parent cloud. The subsequent contraction of the supercritical cores and the formation of stars
within them is dynamic but slower than free-fall. 

A number of authors have recently claimed that 
a series of observational results on the ages of molecular clouds and the timescale of star formation 
suggest that molecular clouds are ``young'' compared to the ambipolar-diffusion timescale, and 
hence these observations 
contradict the predictions of the ambipolar-diffusion
theory (e.g., Hartmann \etal 2001; Elmegreen 2000).
We re-examine these underlying observations and show that the timescales they claim to 
be measuring are severely underestimated, because they only represent {\em part } of 
the relevant processes (star formation and lifetimes of molecular clouds). We then show that the 
quantitative predictions of the ambipolar-diffusion theory for the timescales of the {\em observed}
phases of star formation are in excellent agreement with observations. 
There is no observational support for
 a scenario of ``rapid star formation''.

\section{Age of T-Tauri Stars and its Relation to the
 Star-Formation Timescale and the Age
  of Molecular Clouds \label{ttauri}}

It has been argued that the sound-crossing time in molecular clouds, given their observed 
typical size and temperature,
is $\approx 10^7$ yr and that (1) this places a lower limit on the lifetime of molecular clouds,
and (2) that the clouds are expected to harbor a stellar population of comparable
age (Hartmann \etal 1991). Yet, based on Hayashi-track age estimates, the pre-main--sequence
stars found in surveys are typically younger ($\approx 10^6$ yr)
\citep{Herbig78,CohenK79,Herbig86}. This apparent discrepancy
is known as the ``Post--T-Tauri Problem''.

Hartmann \etal (1991) estimated the age of T-Tauri stars
in the Taurus molecular cloud using pre-main--sequence evolutionary tracks to
obtain their masses and from that the Kelvin-Helmholtz timescale.
They found a relatively short mean age ($\approx {\rm few} \times
10^6$ yr) and they concluded that the
mass estimates from standard Hayashi pre-main-sequence evolutionary tracks are not
reliable. Gomez \etal (1992) repeated the study for a
larger sample of newborn stars at different parts of the Taurus
cloud with the same results. However, this time the conclusion was that
the molecular clouds must be younger than previously thought.
Subsequent work focused on refining the pre-main--sequence
evolutionary tracks, but comparison between different sets of tracks
for low-mass pre-main--sequence stars yield only a factor of 2
difference in the age estimates \citep{White99,Simon01}.
Hence the consensus is that the expected population
of older ($\ge$ 5 Myr) pre-main--sequence stars is missing. 

The above considerations have led to the ``Rapid Star Formation'' 
scenario (Ballesteros-Paredes \etal 1999; Elmegreen 2000; 
Hartmann 2001; Hartmann \etal 2001), which favors short-lived ($\approx$ a few Myr)
molecular clouds and consequently a short ($\approx 1$ Myr) timescale
of star formation. According to this scenario, giant molecular clouds form
rapidly (in a time $\approx 10^6$ yr) due to colliding 
streams of turbulent flows. ``Cores'' (of sizes $\approx 0.1 {\rm
  \,pc}$) within giant molecular clouds then also
form rapidly due to supersonic turbulence and immediately collapse 
dynamically to form stars.

\subsection{Star-Formation Timeline}

Although the observations 
cited above have significantly contributed to a better understanding 
of the T-Tauri phase of star formation, the conclusions that have 
been drawn from them concerning the overall duration of the star-formation 
process and the ages of molecular clouds are not valid. 
They are based on a fundamentally flawed understanding of the star formation timeline. 

%%%%%%%%%%%%%%%%%%%% FIGURE %%%%%%%%%%%%%%%%%%%%%%%
%
\begin{figure}
\plotone{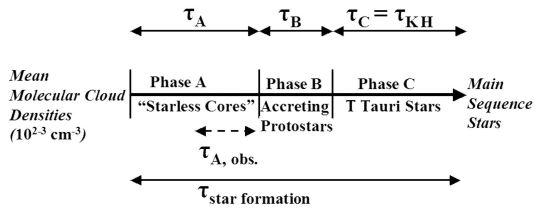}
\caption{\label{fig_timeline}
Star Formation Timeline (not to scale).}

\end{figure}
%%%%%%%%%%%%%%%%%%%%%%%%%%%%%%%%%%%%%%%%%%%%%%%%%%%
To demonstrate this point, we show in Figure \ref{fig_timeline} a
star-formation timeline depicting the star-formation process as a
whole, starting at mean molecular cloud densities and ending with the
emergence of zero-age main-sequence stars. We distinguish
three different phases. In phase A, which has a 
duration $\tau_{\rm A}$, typical ``starless cores'' 
form and contract toward the creation of protostars. 
At the end of phase A, hydrostatic protostellar cores (protostars) have formed,
and accrete mass through phase B, which has a duration $\tau_{\rm B}$. Finally, when mass
accretion stops, the pre-main--sequence stars contract, almost at
constant mass, at the Kelvin-Helmholtz timescale $\tau_{\rm
  KH}$. These objects are known as T-Tauri stars, and the time interval
during which the Kelvin-Helmholtz contraction takes place is phase
C, with duration $\tau_{\rm C}\approx\tau_{\rm KH}$. At the end of phase C, the stars reach
the zero-age main-sequence. 

All the aforementioned observational work estimates the duration of
the T-Tauri phase, or phase C in Figure 1. The ``zero point'' of what the authors
call ``age'' of the T-Tauri stars is the beginning of phase
C. Therefore, all the quoted estimates for the ``ages'' are only
estimates of $\tau_C$. These measurements offer no
information at all on the duration of phases A and B, and hence on the
duration of the star-formation process {\em as a whole}, or on the
ages of molecular clouds. 

The notion that phases A and B are short compared to phase C 
(T-Tauri phase) and therefore the duration of phase C sets the timescale for
the star formation process as a whole {\em is not} (and cannot be) supported by
observations measuring only the duration of phase C. In fact, as we
discuss in \S \ref{sec_stat}, measurements of the duration of phase 
A yield $\tau_{A} \gtrsim \tau_{C}$. Consequently, neglecting
phase A leads to a serious {\em underestimate} of the timescale of
star formation and to erroneous conclusions about the age of molecular clouds.

If the neglected time interval $\tau_A$ corresponding to phase A 
is long compared to the Kelvin-Helmholtz timescale 
(as is typically, but not always, the case in ambipolar-diffusion--induced core formation 
in magnetically subcritical clouds), 
then the overall ages of the pre-main--sequence stars are
$\approx 10^7 $ yr, and the ``post-T-Tauri'' problem no longer exists.

\section{Age Spreads of Stars in Clusters}

Observations of the age spreads of stars in Galactic open clusters have yielded
diverse results. Some observations showed significant
age spreads (e.g., NGC 3293, Herbst \& Miller 1982; NGC 6231, 
Sung \etal 1998), while others found no appreciable 
age spreads (e.g. NGC 6531, Forbes 1996; NGC 3293, Baume \etal 2003). 
Similarly, attempts to use the spread of the estimated ages of T-Tauri stars to
study the history of star-forming regions (Palla \&
Stahler 2000) were hampered by observational errors and uncertainties
(Hartmann 2001). Still, results suggesting small age spreads in star clusters have been 
used  to draw support for the idea that molecular clouds are younger than 
their sound crossing times ($\approx 10^7$ yr) and for the rapid star-formation scenario (e.g., 
Elmegreen 2000).

%%%%%%%%%%%%%%%%%%%% FIGURE %%%%%%%%%%%%%%%%%%%%%%%
%
\begin{figure*}
\plotone{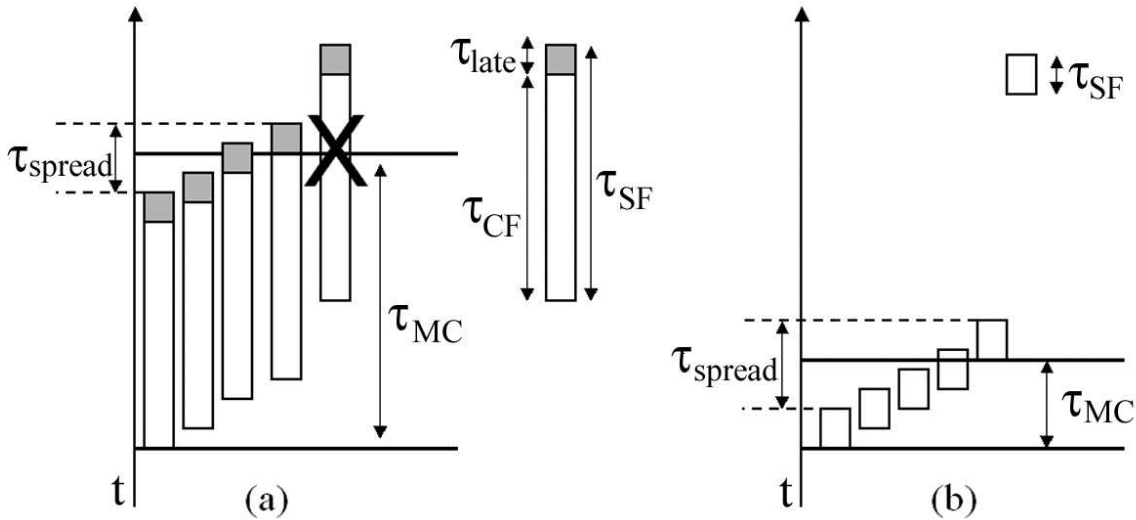}
\caption{\label{fig_spreads}
Interpretation of observations of age spreads in 
star clusters. Time increases upward along the 
vertical axes. The two solid horizontal lines in each panel represent
the instances of birth and dispersion of a molecular cloud.
Each parallelogram represents the process of formation 
of a single star (not to scale). 
(a) Ambipolar Diffusion Theory: the time required for the formation of
a magnetically supercritical core, $\tau_{CF}$, is a large fraction of 
the star-formation timescale, (b) Rapid Star-Formation Scenario: 
the core-formation timescale is neglected {\em by assumption}, 
and the age of the molecular cloud is consequently the same
as the age spread of stars in a stellar cluster within the cloud.}

\end{figure*}
%%%%%%%%%%%%%%%%%%%%%%%%%%%%%%%%%%%%%%%%%%%%%%%%%%%

These conclusions are based on the implicit assumption that
the cluster age spread is always comparable to the molecular cloud lifetime, 
$ \tau_{\rm spread} \approx \tau_{\rm MC}$, which, as shown in Figure
\ref{fig_spreads}, is not true. Figure \ref{fig_spreads} is a sketch of the
star formation history of a molecular cloud according to the ambipolar-diffusion
theory (Fig. \ref{fig_spreads}a), and according to the rapid star-formation scenario
(Fig. \ref{fig_spreads}b). Time increases upward along the 
vertical axes. The two solid horizontal lines represent
the instances of birth and dispersion of the parent cloud.
Each parallelogram represents the time $\tau_{\rm SF}$ 
for the formation of stars from material of mean  
cloud density. The first star forms $\tau_{\rm SF}$
{\em after} the birth of the cloud. In Figure  \ref{fig_spreads}a, 
$\tau_{\rm CF}$ (longer, unshaded part of each parallelogram) 
is the time required for the formation of a magnetically supercritical core 
and is comparable to the ambipolar-diffusion timescale, $\tau_{\rm AD}$; 
and $\tau_{\rm late}$ (shorter, shaded part of each parallelogram)
is the duration of the later stages of star formation.
{\em No stars can form 
after the dispersion of the parent cloud unless a supercritical core has 
already formed by the time of dispersion}. 
Hence, the rightmost parallelogram in Figure \ref{fig_spreads}a is not realizable
in nature and is crossed out in the figure.
Altogether then, if the age spread $\tau_{\rm spread}$ of the stars formed 
during the lifetime of a particular molecular cloud is determined 
observationally, the general relation allowing the determination of the
parent molecular cloud lifetime $\tau_{\rm MC}$ is (see Fig. \ref{fig_spreads}a):
\begin{equation}
\tau_{\rm MC} = \tau_{\rm SF} +\tau_{\rm spread} 
- \tau_{\rm late}\nonumber \\
= 
\tau_{\rm spread} + \tau_{\rm CF} \label{theeq}.
\end{equation}
Using equation (\ref{theeq}), the {\em observed values} of $\tau_{\rm spread}$, and 
the ambipolar-diffusion theory of star formation ( $\tau_{\rm CF}\approx \tau_{\rm AD}$), we
find molecular cloud lifetimes to be consistent with 
molecular cloud age estimates based on cloud crossing times ($\approx 10^7$ yr)
\footnote{This result does not depend on the 
duration of the late stages of star formation, $\tau_{\rm late}$.}.

The interpretation of observations of age spreads according to the 
proponents of the rapid star-formation scenario is shown in 
Figure \ref{fig_spreads}b. In this case, the core formation phase 
is essentially non-existent, $\tau_{\rm CF} \approx 0$, 
so equation (\ref{theeq}) gives $\tau_{\rm MC} = \tau_{\rm spread}$ 
and the molecular clouds are found to be young. However, such an
interpretation {\em cannot} possibly be considered as observational support for the rapid star-formation
scenario because {\em the assumption of rapid star formation, 
$\tau_{\rm CF} \approx 0$, is already built in}.

\section{Statistics of Pre-Stellar Cores and the Timescale of Star Formation}\label{sec_stat}

Another technique for inferring the timescale of star formation has its
origin in the following simple statistical argument: Since a core with
an embedded point source (hydrostatic protostellar object) is preceded,
in an evolutionary sense, by a starless core, then the ratio of the number of cores
with embedded point sources and the number of starless cores should
reflect the ratio of the duration of each phase; i.e., 
\begin{equation}
\frac{\tau_{\rm starless \, core}}{\tau _{\rm embedded \, p.s.}}=
\frac{\rm \# \,\, starless \,\,cores}{\rm \# \,\,cores \,\,with
  \,\,embedded \,\, point \,\,sources}
\,
\end{equation}
(Beichman \etal 1986).
This method has been employed to estimate the 
evolutionary timescale of prestellar cores (Ward-Thompson \etal 1994; 
Jijina \etal 1999;  Lee \& Myers 1999; Jessop \& 
Ward-Thompson 2000). 

Lee \& Myers (1999) presented an all-sky survey of optically selected
cores and found an overall ratio of cores with embedded point sources
and starless cores of about 0.3. Using values for the evolutionary
timescale of the embedded point-source phase  
between $1-5 \times 10^5$ yr (Class 0 and Class I
objects), they concluded that the
duration of the starless core phase is about $10^6$ years. Based on a survey of 
NH$_3$ cores, Jijina \etal (1999) found a similar result. 
In both cases, the authors concluded that the number of detected
prestellar cores is too small to be consistent with
ambipolar-diffusion--initiated star formation predictions. 

Ward-Thomson \etal (1994) and Jessop \& Ward-Thompson (2000) used 100
${\rm \mu m}$ and 60 ${\rm \mu m}$ all sky surveys and found 
a ratio $\approx 1/10$ of cores with embedded point sources and starless cores.
From this result, using a value $\tau _{\rm embedded \, p.s.} = 10^6$
years (including in this phase Class 0, Class I, and Class II objects),
they found $\tau_{\rm starless \, core} = 10^7$ years and
concluded that this value is consistent with the ambipolar-diffusion theory of the
starless core phase.

Caution should be exercised when using the results obtained in the
way described above.
Observationally, only the ratio of $\tau_{\rm starless \, core}$
and  $\tau _{\rm embedded \, p.s.}$ can be determined. A specific
estimate for either one of the two timescales requires a theoretical
value for the other. Any theoretical uncertainty in the 
assumed timescale is carried over to the estimated one. 
However, even if one accepts the value for  $\tau_{\rm starless \,
  core}$ as derived from an assumed $\tau _{\rm embedded \, p.s.}$,
one should be very careful in
 comparing it with ambipolar-diffusion--induced core formation and evolution timescales.
First, typical starless cores spend an appreciable
fraction of phase A in  Figure \ref{fig_timeline} having column-density contrasts with the parent
cloud too low to be identified 
as ``cores''. Second, {\em the actual duration of the
starless core phase is not a universal number in the ambipolar-diffusion theory. It depends
on the mass-to-flux ratio} ($M/\Phi_{B}$) {\em of the parent molecular cloud} \citep{FM93, CB01}.

%%%%%%%%%%%%%%%%%%%% FIGURE %%%%%%%%%%%%%%%%%%%%%%%
%
\begin{figure}
\plotone{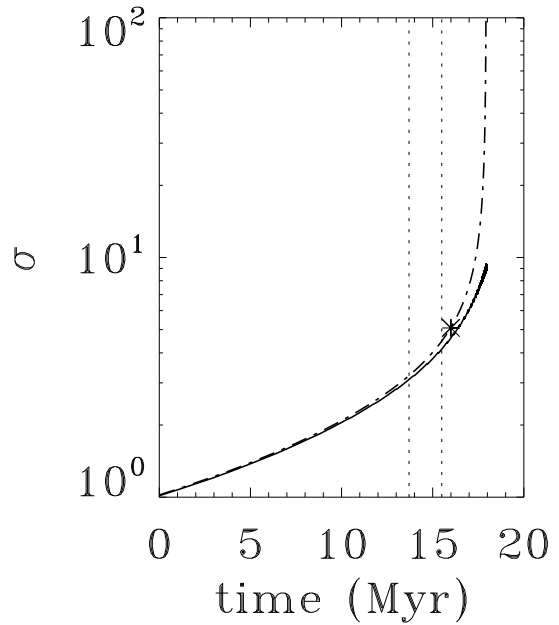}
\caption{\label{fig_column_a} Column density $\sigma$ 
(in units of the mean column density of the parent cloud) of an 
ambipolar-diffusion--controlled fragment (or ``core'') as a function of
time. The initial mass-to-flux ratio (in units of its critical value for
collapse) is $\mu_{0} = 0.25$. {\em Solid line}: average $\sigma$ when fragment is viewed
``face-on''. {\em Dot-dashed line}: $\sigma$ at the center of the fragment.
The dotted lines mark the instances at which column-density enhancements
of 3 and 4 are achieved, with respect to the initial (parent-cloud) column density.}
\end{figure}
%%%%%%%%%%%%%%%%%%%%%%%%%%%%%%%%%%%%%%%%%%%%%%%%%%

To demonstrate the first point, we plot in Figure \ref{fig_column_a} the column density
of a model ambipolar-diffusion--controlled ``core'',
in units of the background column density of the parent
cloud, as a function of time. The solid line represents the value of
the average column density of the oblate fragment when observed face-on out to 
a radius where a magnetically supercritical core eventually
forms. The dot-dashed line corresponds to the value of the central
column density of the core.

In all the observational work reviewed above, 
a density contrast of at least a factor 2 - 4 is required in order to  
identify an object as a ``core''. As seen clearly in Figure \ref{fig_column_a}, the model cloud
requires $\approx 10^7$ years before it reaches a
column-density contrast of 2 with respect to the background (parent cloud), during which
time it would not be identified as a core in observational
surveys. A density contrast of a factor of 3 is not developed until
after about $1.4 \times 10^7$ years. Hence, 
observational surveys miss a large part of the evolutionary phase
preceding the formation of a supercritical core, 
and the actual duration of phase A 
is therefore severely underestimated; i.e., the fraction of phase A observable by these surveys is only a small
fraction of the entire duration of phase A
($\tau_{\rm A, \, obs} \ll \tau_{\rm A}$ of Fig. \ref{fig_timeline}).

%%%%%%%%%%%%%%%%%%%% FIGURE %%%%%%%%%%%%%%%%%%%%%%%
%
\begin{figure}
\plotone{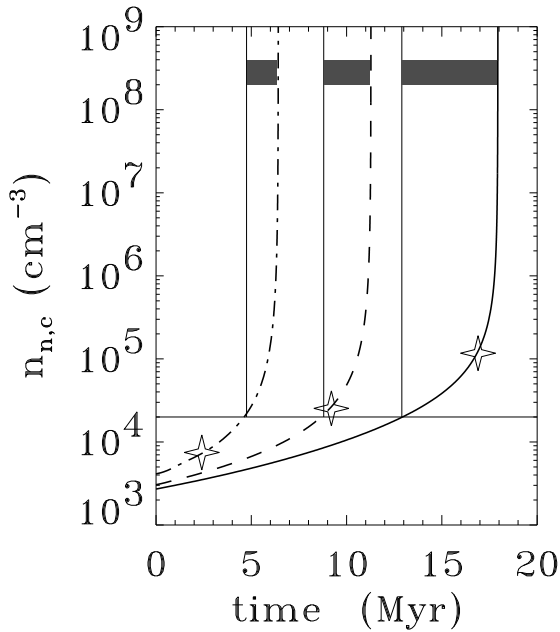}
\caption{\label{fig_column_b} Time evolution of the central number
density of a core for three different values of the initial
mass-to-flux ratio $\mu_0= (M/\Phi_B)/(M/\Phi_{B})_{\rm
  critical}$: $\mu_0=0.25$ ({\em solid line}), $\mu_0=0.5$ ({\em dashed line}), and
$\mu_0=0.8$ ({\em dot-dashed line}). The horizontal line represents the
central density of a well defined ammonia core, $2\times 10^4$
cm$^{-3}$, while the ``star'' on each curve marks the time of formation of a
supercritical core in each case ($\approx$ 17, 9, and 2 Myr, respectively).}
\end{figure}
%%%%%%%%%%%%%%%%%%%%%%%%%%%%%%%%%%%%%%%%%%%%%%%%%%%%

The model cloud of Figure \ref{fig_column_a} is characterized by a relatively small initial
mass-to-flux ratio [$\mu_0= (M/\Phi_B)/(M/\Phi_{B})_{\rm critical} =$ 0.25]. To demonstrate the
dependence of the duration of phase A on $\mu_0$, we plot in Figure \ref{fig_column_b} the time evolution of the
central number density of the core for three different values of
$\mu_0$ (see, also, Fiedler \& Mouschovias 1993, Fig. 9a). The horizontal solid line corresponds to the minimum central
density of a well-defined ammonia core ($2 \times 10^4 \, {\rm
  cm^{-3}}$). The shaded bands represent the duration of the
observable (using ammonia cores) part of phase A. For the most
subcritical cloud ($\mu_0=0.25$, {\em solid line}), $\tau_{\rm A, obs}
\approx 5 $ Myr, while it is only $\approx 2$ Myr for the case $\mu_0=0.5$ 
({\em dashed line}), and $\approx 1$ Myr for the case $\mu_0=0.8$
({\em dot-dashed line}). Thus the lifetime
of the observable ``starless core'' predicted by the ambipolar-diffusion theory depends on the value of the 
mass-to-flux ratio of the {\em parent} molecular cloud, and can be $<1$ Myr, in complete agreement
with observations. The closer to the critical value the mass-to-flux ratio is, the smaller the observable 
starless-core lifetime, with a lower limit being the magnetically diluted free-fall timescale.

\section{Conclusions}
 
We have re-examined the observational estimates of the lifetimes of 
molecular clouds and the timescale of star formation (based on ages of T-Tauri stars, 
age spreads of stars in clusters, and statistics of pre-stellar cores).
{\em All available measurements or estimates of the duration 
of different phases of the star formation process and of molecular-cloud lifetimes 
are completely consistent with the theoretical predictions of the ambipolar-diffusion--controlled 
fragmentation of molecular clouds and evolution of the protostellar fragments}.

Observations of ages of T-Tauri stars only measure the duration of the late stages 
of star formation (after the formation of a hydrostatic core) and offer no information on 
the earlier stages, which can be (and, according to the ambipolar-diffusion
theory, typically are) of much longer duration. These observations are consistent 
with molecular-cloud lifetimes $\approx 10$ Myr.

Observations of age spreads of stars in clusters can only be used to derive molecular-cloud
lifetimes under some assumption concerning the core formation timescale $\tau_{\rm CF}$
(see Eq. [\ref{theeq}]).
In the ambipolar-diffusion--induced star formation theory, $\tau_{\rm CF}$ is the 
ambipolar-diffusion timescale, and observations suggesting 
$\tau_{\rm spread} \approx 1{\rm \, Myr}$ are consistent with molecular-cloud 
lifetimes of $\approx 10 {\rm \, Myr}$. 

In a typical molecular cloud, ambipolar diffusion
requires $\approx 10 {\rm \, Myr}$ to produce a core
with a factor of 2 density contrast with respect to the parent cloud. It is only after
this time that the object can be identified as a core in observational surveys.
The remaining, observable lifetime of a ``starless core'' is only a few Myr, a value consistent with 
observations measuring the ratio of starless-core and embedded-point--source lifetimes.
The observable lifetime of a starless core depends, according to the ambipolar-diffusion 
theory, on the initial mass-to-flux ratio of the parent cloud. The closer to
critical this ratio is, the shorter the observable ``starless core'' lifetime 
(which can be $\lesssim 1 {\rm \, Myr}$ for nearly-critical clouds).

Altogether, there is no observational evidence suggesting that molecular clouds
are younger than their sound-crossing times. Therefore, ``rapid'' star formation is
{\em not} an ``observational requirement'', all statements to the contrary notwithstanding.

\acknowledgements{ We thank C. Lada and  V. Pavlidou for useful discussions. 
This work was carried out without external support, and would
not have been published without the generosity of the ApJ.}

%%%%%%%%%%%%%%%%%%%%%%%%%%%%%%%%%%%%%%%%%%%%%%%%%%%%%%%
% end of main text
%%%%%%%%%%%%%%%%%%%%%%%%%%%%%%%%%%%%%%%%%%%%%%%%%%%%%%%

\end{document}